\newif\ifCAMREADY{}

\ifCAMREADY{}
\documentclass[sigconf]{acmart}
\else
\documentclass[sigconf,nonacm=true]{acmart}
\fi

\author{Billy Bob Brumley}
\orcid{0000-0001-9160-0463}
\affiliation{
\institution{Rochester Institute of Technology}
\city{Rochester, NY}
\country{USA}
}
\email{bbbics@rit.edu}

\usepackage[T1]{fontenc}
\usepackage{graphicx}
\usepackage{xspace}
\usepackage{lipsum}
\usepackage[vlined,figure]{algorithm2e}

\newcommand{\PARAGRAPH}[2][.]{\subsubsection*{#2#1}}
\newcommand{\FR}{Flu\-sh+\allowbreak{}Re\-load\xspace}
\newcommand{\MCHAMMER}{MC\-Ham\-mer\xspace}

\newcommand{\KEYWORDS}{%
Side-channel analysis,
microarchitectural timing attacks,
machine clears,
self-modifying code,
\FR{},
\MCHAMMER{},
leakage assessment}

\title{U~Can~Touch~This! Microarchitectural~Timing~Attacks~via~Machine~Clears}

\begin{abstract}
Microarchitectural timing attacks exploit subtle timing variations caused by
hardware behaviors to leak sensitive information. In this paper, we introduce
\MCHAMMER{}, a novel side-channel technique that leverages machine clears
induced by self-modifying code detection mechanisms. Unlike most traditional
techniques, \MCHAMMER{} does not require memory access or waiting
periods, making it highly efficient.
We compare \MCHAMMER{} to the classical \FR{}
technique, improving in terms of trace granularity,
providing a powerful side-channel attack vector.
Using \MCHAMMER{}, we successfully recover keys
from a deployed implementation of a cryptographic tool.
Our findings highlight the practical implications
of \MCHAMMER{} and its potential impact on real-world systems.
 \end{abstract}

\ifCAMREADY{}
\copyrightyear{2025}
\acmYear{2025}
\setcopyright{cc}
\setcctype{BY-NC-SA}
\acmConference[SAC '25]{The 40th ACM/SIGAPP Symposium on Applied Computing}{March 31--April 4, 2025}{Catania, Italy}
\acmBooktitle{The 40th ACM/SIGAPP Symposium on Applied Computing (SAC '25), March 31--April 4, 2025, Catania, Italy}
\acmDOI{10.1145/3672608.3707711}
\acmISBN{979-8-4007-0629-5/25/03}

\fi

\begin{document}

\maketitle

\section{Introduction}\label{sec:intro}

Side-channel attacks pose a significant threat to security-critical
systems, exploiting unintended information leakage through physical or
microarchitectural channels. Over the past two decades, researchers have developed
various techniques to capture and analyze these side channels,
where the leakage source fundamentally varies:
examples include
remote timings that vary due to data caching \cite{2005:bernstein},
Prime+Probe that can target data at different cache levels \cite{DBLP:conf/ctrsa/OsvikST06},
L1 data-cache timings on Simultaneous multithreading (SMT) architectures \cite{Percival05},
L1 instruction-cache timings on SMT architectures \cite{DBLP:conf/ches/AciicmezBG10},
branch prediction analysis (BPA) that concerns the branch target buffer \cite{DBLP:conf/ctrsa/AciicmezKS07},
\FR{} that targets shared memory \cite{2014:fr},
CacheBleed that extracts inter-cache line leakage on SMT architectures \cite{DBLP:conf/ches/YaromGH16},
TLBleed that targets virtual memory translations on SMT architectures \cite{DBLP:conf/uss/GrasRBG18},
and
port contention on SMT architectures \cite{DBLP:conf/sp/AldayaBHGT19}.
Although out of scope for this paper,
these low-level gadgets often build the foundation for speculative and transient execution attacks:
for example,
\FR{} for Meltdown \cite{DBLP:conf/uss/Lipp0G0HFHMKGYH18}
and BPA for Spectre \cite{DBLP:conf/sp/KocherHFGGHHLM019}.

Despite their
effectiveness, these traditional techniques often face limitations in
granularity and detection capabilities, particularly in high-performance
environments. This paper introduces \MCHAMMER{}, a novel side-channel analysis
technique designed to enhance leakage and information extraction.

\MCHAMMER{} leverages the concept of machine clears---specific microarchitectural
events that disrupt normal execution flow---to induce measurable timing variations
in code traversal. Unlike traditional cache-based attacks, which rely
on memory access patterns and cache hits/misses, \MCHAMMER{} exploits the latency
variations caused by these machine clears. This approach not only increases the
granularity of the resulting side-channel signals, but improves accuracy.

To quantitatively assess the effectiveness of \MCHAMMER{}, we utilize the
Normalized Inner Class Variance (NICV) metric. NICV is particularly advantageous
because it depends solely on public data and remains independent of specific
leakage models, making it a versatile tool for comparing the quality of
different trace datasets. NICV facilitates
a direct comparison between the \FR{} and \MCHAMMER{} methods, as
illustrated in our experimental results.

Our findings demonstrate that \MCHAMMER{} significantly outperforms \FR{},
with an increase in trace granularity by a factor ranging from 13 to 15 times.
This enhanced visibility translates into
better detection of critical leakage points when targeting cryptographic operations, thereby enabling
more efficient and controlled side-channel attacks.

We apply the \MCHAMMER{} technique to a practical cryptographic scenario involving
the Elliptic Curve Digital Signature Algorithm (ECDSA). By targeting the SECCURE
cryptographic utility and monitoring the scalar multiplication process, we
successfully recover the secret nonce scalar from a single trace, and thus the
private key. This result underscores \MCHAMMER{}'s potential as a powerful tool for
side-channel analysis, capable of revealing the state of security-critical
cryptographic algorithms with minimal traces and high accuracy.

In conclusion, \MCHAMMER{} represents a significant advancement in side-channel
attack methodologies, offering higher granularity and improved trace
accuracy. Its application to cryptanalytic key-recovery attacks demonstrates its
practical utility and effectiveness in real-world environments.
The remainder of this paper is organized as follows.
\autoref{sec:background} reviews necessary background to understand the nature of \MCHAMMER{}.
\autoref{sec:leak} revisits the \FR{} method to help unveil this new attack vector, and we present the \MCHAMMER{} gadget.
\autoref{sec:covert} compares timing distributions between the \FR{} and \MCHAMMER{} techniques, outlining an abstract covert channel enabled by the machine clear latency.
\autoref{sec:side} compares granularity and accuracy between the \FR{} and \MCHAMMER{} techniques, leveraging the NICV metric to quantify our improvement.
\autoref{sec:attack} details the \MCHAMMER{}-based
key-recovery attack on a deployed, open-source implementation of a digital signature algorithm.
\autoref{sec:related} reviews specific work that is closely related to the \MCHAMMER{} technique
in the field of side-channel attacks.
We close in \autoref{sec:conclusion}.
\section{Background}\label{sec:background}

\subsection{The \FR{} Technique}
\FR{} is a side-channel attack technique used to exploit timing
differences in microarchitectural behavior of modern CPUs to leak sensitive
information \cite{2014:fr}.
The technique leverages the shared memory and cache hierarchy in systems where
an attacker can access the same physical memory locations as the victim. This
typically occurs in environments like virtual machines, shared libraries, or
cloud services. The technique exploits the timing differences between accessing
data from the cache versus main memory.

\begin{algorithm}
\DontPrintSemicolon{}
\KwIn{Memory Address \textit{addr}.}
\KwResult{True if the victim accessed the address.}
\SetKwBlock{Begin}{begin}{end}
\SetKwData{CurrentTIme}{CurrentTime}
\BlankLine{}

\Begin{%
  \texttt{flush}\null(\textit{addr})\;
  \texttt{Wait} \textit{for the victim.}\;
  time \(\gets\) \texttt{current\_time\null()}\;
  tmp \(\gets\) \texttt{read}\null(\textit{addr})\;
  readTime \(\gets\) \texttt{current\_time\null()} - time\;
  \Return{} readTime \(<\) threshold\;
}

\caption{The \FR{} microarchitectural timing attack technique.}\label{alg:fl}
\end{algorithm}

As \autoref{alg:fl} illustrates, one iteration of the \FR{} technique consists of three steps.
\begin{enumerate}
\item The attacker flushes a specific cache line by issuing the \texttt{clflush}
      instruction, which ensures that the targeted data is removed from all levels
      of the cache hierarchy.
\item The attacker waits for the victim to potentially access the data. During
      this time, the CPU may load the data back into the cache if the victim uses it.
\item The attacker accesses the same memory location again and measures the
      access time. If the data is still in the cache (indicating the victim accessed
      it), the access will be fast. If the data is not in the cache, the access will
      be slower due to a main memory fetch.
\end{enumerate}

When data is in the cache, access times are significantly lower compared to when
data must be fetched from main memory. By repeatedly flushing, waiting, and
reloading, the attacker can determine which cache lines have been accessed by
the victim.

\subsection{Machine Clears}
A machine clear on Intel architectures refers to the processor's internal
mechanism for handling certain exceptional conditions that require flushing or
restarting the pipeline to maintain correct execution. These conditions cause
the processor to clear or invalidate its pipeline and start over, ensuring the
correct execution of instructions despite disruptions.

There are several types of machine clears mentioned in the Intel optimization manual \cite[Table 22-1]{2024:intel},
each triggered by specific conditions; here are four common types.

\begin{enumerate}
\item When a program modifies its own executable code while it is running, this causes a Self-Modifying Code (SMC) machine clear.
      The processor needs to ensure that it is not executing outdated or incorrect instructions that may have been modified.
      The pipeline is flushed to discard any prefetched or partially executed instructions, and the modified code is re-fetched and executed from memory.

\item Certain floating-point operations, such as exceptions or conditions that require special handling, cause a Floating Point (FP) machine clear.
      The pipeline is cleared to handle the floating-point exception or special condition correctly.
      This ensures that the floating-point operations are executed accurately and in accordance with the required precision and exception handling rules.

\item When the processor detects a potential violation of memory ordering rules,
      which dictate the order in which loads and stores to memory must occur to ensure
      correct program execution, this causes a Memory Ordering machine clear.
      The pipeline is cleared to enforce the correct order of memory operations.
      This ensures that memory consistency is maintained, and all operations are executed in the correct sequence, preserving data integrity.

\item When the processor misinterprets the dependencies between memory accesses,
      particularly when it incorrectly assumes that two memory operations are
      independent (disambiguation) but later detects a conflict, this causes a Memory Disambiguation machine clear.
      The pipeline is flushed to resolve the detected memory dependency conflict.
      This clear allows the processor to re-execute the instructions with the correct understanding of their dependencies, ensuring correct program execution.
\end{enumerate}

To summarize, these machine clears are crucial for maintaining the reliability
and correctness of instruction execution on Intel processors. They handle
specific conditions where the pipeline must be restarted to correct or prevent
errors.
\section{\MCHAMMER{}: Concept}\label{sec:leak}

In this section, we take a wider look at the microarchitectural impact of the
\FR{} technique, beyond caching effects, in an attempt to discover other
contributions to the root cause and other microarchitectural attack vectors. Our
analysis leads us to develop the \MCHAMMER{} gadget, which we compare and contrast
with the \FR{} technique.

\PARAGRAPH{Environment}%
We performed all the experiments in this paper on an
Intel Core i5-7500T (Kaby Lake) CPU clocked at 2.70GHz.
The CPU has four physical cores and does not feature HyperThreading (HT, SMT).
The system is equipped with 16GB of DDR4 and is running Debian 12 (bookworm).
We disabled TurboBoost and set the CPU frequency governor to \texttt{performance}.

\subsection{\FR{}: Revisited}

Our investigation began by examining the \FR{} microarchitectural
technique, specifically in the context of code residing in executable pages.
Recall that with this technique, the attacker flushes a specific cache
line, effectively removing it from all levels of cache hierarchy.
When the victim process subsequently accesses this data, it results
in a cache miss, causing the data to be reloaded into the cache. The
attacker then measures the time taken to access the same cache line. If
the data was reloaded by the victim, the attacker experiences a cache hit,
revealing that the victim accessed the same data.

Hence, cache hits and misses are fundamental to the \FR{} technique. In
the context of code, the victim accessing the data typically means the code being
executed, while the attacker accessing the data is usually through a \verb+mov+ instruction.
So while the victim will experience instruction cache hits and misses,
there could potentially be many microarchitectural effects in play---not
solely the instruction cache.

\begin{figure}[!t]
  \centering
  \includegraphics[width=\linewidth]{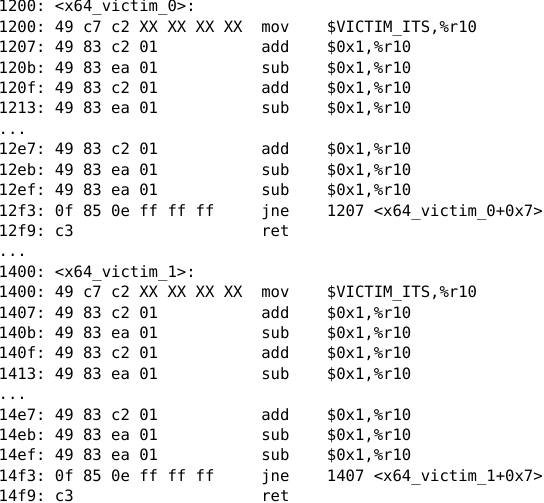}
  \caption{A strawman shared library utilized by the victim (from \texttt{objdump} output).}%
  \label{fig:victim}
\end{figure}

To investigate these code aspects further,
we developed a strawman shared library to serve as a victim,
illustrated in \autoref{fig:victim}.
The library contains two functions: \texttt{x64\_\-victim\_\-0} and \texttt{x64\_\-victim\_\-1}.
These two functions have identical contents, just aligned at different boundaries.
Their entry points are 512 bytes apart, and each function consists of 256 bytes,
or four 64-byte instruction cache lines.
Each function enters what is---for our intents and purposes---an infinite loop,
where the body of the loop simply increments and decrements a counter in a serial fashion.
We set the number of loop iterations (\texttt{VICTIM\_\-ITS})
to \(2^{30}\) for the experiments in this section.

Linking against this shared library, we ran two experiments.
In the first experiment, the victim calls \texttt{x64\_\-victim\_\-0}.
In parallel and on a different core,
the attacker mounts a \FR{} attack,
with a probe targeting 128 bytes within the \texttt{x64\_\-victim\_\-0} entry point.
The attacker collects \(2^{20}\) latency measurements.
We configured the probe to use a 1000-cycle waiting period,
which is the smallest reasonable value \cite[Table~2]{2016:degrade}.
Lower waiting periods improve \FR{} trace granularity,
at the expense of potentially introducing noise.
The waiting period is typically very dependent on execution characteristics
of the victim.
Generally speaking, the faster and more frequently a probed function executes,
the shorter the waiting period needs to be.
The second experiment is otherwise identical,
but with the victim calling \texttt{x64\_\-victim\_\-1},
while the attacker keeps the original probe location
(i.e., 128 bytes within the \texttt{x64\_\-victim\_\-0} entry point).

The only metric considered in the original \FR{} work are
these empirical latencies \cite[Section~3]{2014:fr}.
Instead of empirical latencies, we chose to gather metrics using the
Performance Monitor Unit (PMU).

The PMU on Intel CPUs is a hardware feature designed
to collect and report various performance-related data, such as CPU cycles,
instructions executed, cache hits and misses, branch predictions, and other
low-level system events. This data is critical for performance analysis and
optimization. The Linux \verb+perf+ tool interfaces with the PMU by leveraging the
kernel's performance monitoring capabilities, allowing users to set up and
manage hardware performance counters, collect data, and analyze performance
metrics.

In the context of studying the \FR{} technique,
one of the most useful PMU counters available is \texttt{L1\--icache\--load\--misses},
which tracks the number of L1 instruction cache misses.
Reading the software manual,
we also identified the \texttt{machine\_clears.smc} PMU counter
as a potentially interesting metric, which tracks the number of machine clears
induced by Self-Modifying Code (SMC) detection.
Regarding the two \FR{} experiments previously discussed,
we executed the attacker code in both cases with \verb+perf+
to collect the values of both of these PMU counters
(\texttt{L1\--icache\--load\--misses}, \texttt{machine\_clears.smc}).
\autoref{tab:perf1} contains the results.

\begin{table}[!b]
  \caption{Metrics from \texttt{perf} for the \FR{} (FR) experiments.}%
  \label{tab:perf1}
  \centering
  \resizebox{1.0\linewidth}{!}{%
      \begin{tabular}{lrr} \hline
                                    & \texttt{L1\--icache\--load\--misses} & \texttt{machine\_clears.smc} \\ \hline
        FR, \texttt{x64\_\-victim\_\-0} & 1,170,621                      & 1,048,510 \\
        FR, \texttt{x64\_\-victim\_\-1} & 119,354                        & 345 \\ \hline
      \end{tabular}
  }
\end{table}

We first discuss the results regarding the \texttt{L1\--icache\--load\--mis\-ses} metric.
When the victim executes \texttt{x64\_\-victim\_\-0},
the empirical value 1,170,621 is consistent
with the expectation, given that the attacker's \FR{}
probe---set within the \texttt{x64\_\-victim\_\-0} function---collects
\(2^{20}\) measurements.
Roughly each time the victim performs an iteration of the loop,
it will encounter a single cache miss, induced by the attacker issuing
\texttt{clflush} instructions targeted within said loop.
In contrast, when the victim executes \texttt{x64\_\-victim\_\-1},
it experiences significantly fewer L1 instruction cache misses (119,354).
Since the attacker probe remains fixed within the
\texttt{x64\_\-victim\_\-0} function, the attacker issuing \texttt{clflush}
instructions targeted within the \texttt{x64\_\-victim\_\-0} has little
effect on the victim executing within the \texttt{x64\_\-victim\_\-1} function.
To summarize, the \texttt{L1\--icache\--load\--misses} metric seems consistent
with the expectation from the \FR{} experiments.

The surprising result from these two experiments concerns
the \texttt{machine\_clears.smc} metric.
When the victim executes \texttt{x64\_\-victim\_\-0},
the empirical value 1,048,510 is very close to the number of
L1 instruction cache misses (1,170,621).
Roughly each time the victim performs an iteration of the loop,
it passes through code recently (or soon-to-be) targeted by
the attacker's \texttt{clflush} instructions.
This traversal and \texttt{clflush} combination seems to be triggering
the CPU's SMC detection mechanisms, inducing a machine clear.
In contrast, when the victim executes \texttt{x64\_\-victim\_\-1},
it experiences a negligible number of machine clears (345).
This indeed leads us to conclude that the \texttt{clflush}
instructions in combination with traversal of the targeted code
is the root cause of the machine clears.
Since the attacker probe remains fixed within the
\texttt{x64\_\-victim\_\-0} function, and the victim---confined
within the \texttt{x64\_\-victim\_\-1} function---never passes
through those specific \texttt{clflush} addresses,
the victim does not experience any significant number of
machine clears.

To summarize, applied to these two experiments,
the \FR{} technique focuses on exploiting microarchitectural
effects that relate to the \texttt{L1\--icache\--load\--misses} metric;
we now examine how to exploit microarchitectural effects that relate to
the \texttt{machine\_\-clears.\-smc} metric.

\subsection{The \MCHAMMER{} Gadget}

In the context of code execution,
the \FR{} technique requires memory reads and waiting periods
by the attacker, and relies on latencies of the L1 instruction cache;
in contrast, the data in \autoref{tab:perf1} suggests that it is
feasible to construct a side channel without reads, waiting periods,
or direct reliance on L1 instruction cache latencies, but rather
the latency of machine clears induced by victims traversing
frequently flushed code.

\begin{figure}[!t]
  \centering
  \includegraphics[width=\linewidth]{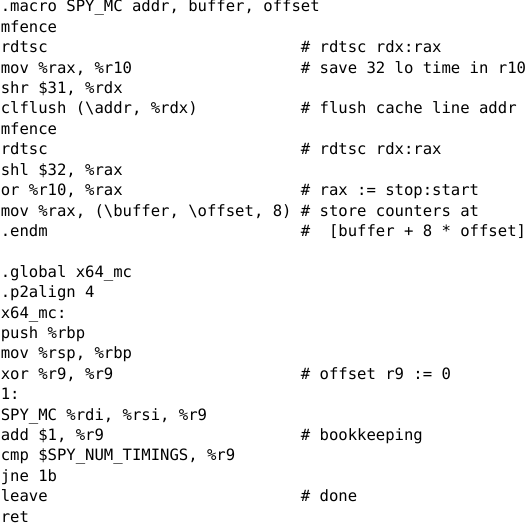}
  \caption{The \MCHAMMER{} gadget for microarchitectural side-channel analysis.}%
  \label{fig:mchammer}
\end{figure}

To that end, \autoref{fig:mchammer} is our proposed microarchitectural attack
gadget that exploits the latency caused by machine clears that are triggered by
SMC-detection mechanisms while flushing code from the cache that is actively being executed.
In contrast to the three-step approach of \FR{}---flush,
wait, reload---\MCHAMMER{} is essentially a single-step gadget built around issuing
the \texttt{clflush} instruction. At a high level, the concept of \MCHAMMER{} is to
time the latency of a single cache flush instruction. If the latency is low,
that indicates the absence of a machine clear, and the victim is not executing
at that address; if the latency is high, that indicates an induced machine
clear, and the victim is indeed executing at that address.

To implement this \MCHAMMER{} concept,
focusing on the macro \texttt{SPY\_MC} in \autoref{fig:mchammer}, the first
\texttt{rdtsc} instruction establishes the start time. The gadget then saves the
low 32 bits to a separate register, to avoid subsequent clobbering. The
\texttt{shr} instruction saves the high bit of the 64-bit \texttt{rdtsc} output;
it overwhelmingly takes a value of zero. The purpose is to prevent the CPU from
reordering the preceding \texttt{rdtsc} instruction, ensuring the
\texttt{clflush} instruction executes strictly after the \texttt{rdtsc}
instruction; i.e., it sets up an artificial dependency since the \texttt{clflush}
instructions takes the \texttt{rdtsc} output as input\footnote{It would be
possible to instead use barriers such as \texttt{lfence} to prevent reordering;
however, barriers are costly in terms of CPU cycles.}.
The base address for the \texttt{clflush} instruction is the virtual address of
the victim code. The second and final \texttt{rdtsc} instruction establishes the
end time. The gadget stores the discrete counter values rather than their
difference (latency), which is useful for understanding granularity and temporal
issues and detecting external effects such as OS ticks and preemptions. The
\texttt{mfence} instructions wrap the \texttt{rdtsc} instruction, acting as a
barrier to prevent aggressive reordering and contiguous measurements from
interfering with each other.

Armed with our new \MCHAMMER{} gadget, our next task was to rerun the two
previous experiments involving the \autoref{fig:victim} shared library.
On the victim side, absolutely nothing in the experiment changed.
In the first experiment, the victim executes solely within the
\texttt{x64\_\-victim\_\-0} function, and in the second experiment, solely within
the \texttt{x64\_\-victim\_\-1} function. The attacker executes the \MCHAMMER{} code
from \autoref{fig:mchammer} in parallel during both experiments, setting the
number of measurements (\texttt{SPY\_\-NUM\_\-TIMINGS}) similarly to \(2^{20}\).
In both experiments, the target address is the same as that which was used in
the \FR{} executions---i.e., 128 bytes within the \texttt{x64\_\-victim\_\-0}
function's entry point. And recall, this target address for the \texttt{clflush}
instruction remains the same in both experiments\@. \autoref{tab:perf2} contains
the results, juxtaposed with the previous \autoref{tab:perf1} experiment results
for clarity.

\begin{table}[!b]
  \caption{Metrics from \texttt{perf} for the \FR{} (FR) and \MCHAMMER{} (MC) experiments.}%
  \label{tab:perf2}
  \centering
  \resizebox{1.0\linewidth}{!}{%
      \begin{tabular}{lrr} \hline
                                    & \texttt{L1\--icache\--load\--misses} & \texttt{machine\_clears.smc} \\ \hline
        FR, \texttt{x64\_\-victim\_\-0} & 1,170,621                      & 1,048,510 \\
        FR, \texttt{x64\_\-victim\_\-1} & 119,354                        & 345 \\ \hline
        MC, \texttt{x64\_\-victim\_\-0} & 1,127,478                      & 1,011,474 \\
        MC, \texttt{x64\_\-victim\_\-1} & 110,115                        & 390 \\ \hline
      \end{tabular}
  }
\end{table}

The results from these new experiments utilizing our new \MCHAMMER{} gadget are
strikingly similar to the \autoref{tab:perf1} results that utilize the
\FR{} technique. With the \texttt{L1\--icache\--load\--misses} metric, we can
see that the victim experiences roughly \(2^{20}\) L1 instruction cache misses
(1,127,478) due to the \MCHAMMER{} gadget when executing within the
\texttt{x64\_\-victim\_\-0} function, and this is virtually the same as the result
with the \FR{} technique (1,170,621). Moving to the
\texttt{x64\_\-victim\_\-1} function, the \MCHAMMER{} (110,115) and \FR{}
(119,354) numbers are low and also consistent. This consistency in the case of
both functions suggests that the impact on the L1 instruction cache is similar
with both methods.

Turning to the \texttt{machine\_clears.smc} metric, we can see that the victim
experiences roughly \(2^{20}\) machine clears (1,011,474) due to the \MCHAMMER{}
gadget when executing within the \texttt{x64\_\-victim\_\-0} function, and this is
virtually the same as the result with the \FR{} technique (1,048,510).
Moving to the \texttt{x64\_\-victim\_\-1} function, the \MCHAMMER{} (390) and
\FR{} (345) numbers are negligible. This consistency in the case of both
functions suggests that the impact on the number of machine clears is similar
with both methods.

To summarize, our new \MCHAMMER{} gadget is able to obtain the same results as the
\FR{} technique regarding these two experiments targeting code, yet it
obtains said results without any memory access nor waiting period. This fact
will be tremendously useful when utilizing our new \MCHAMMER{} gadget as a
side-channel technique, later in \autoref{sec:side}.
We found the \texttt{machine\_clears.smc} counter documented \cite[Table 11]{2017:intel}
as far back as Nehalem (2008), suggesting the \MCHAMMER{} gadget applies across
a wide range of Intel chips.
\section{\MCHAMMER{}: Covert Channel}\label{sec:covert}

In the previous section, we considered only metrics provided by the Linux
\texttt{perf} utility that interfaces with the CPU's PMU\@.
While that is convenient for proving the concept of \MCHAMMER{},
in reality, userspace attackers will not have access to the PMU\@.
In this section, we shift to empirical timings,
and use these latencies to construct a covert channel.
This will eventually lead organically to a side channel later in \autoref{sec:side}.
Note that \MCHAMMER{} inherits \FR{}'s read-only shared memory requirement;
in this section, we realize that by using a shared library,
while \autoref{sec:side} instead uses the (unprivileged) \texttt{mmap} function
to share a system-wide read-only utility (executable).

\subsection{Latencies}

From a metric perspective, the crux of the \FR{} technique is
that the latency between a cache hit and a cache miss typically differs by a few hundred clock cycles.
Without any victim access, the default behavior of the \FR{} technique will be simply a series of cache misses, i.e., high latency.
That is due to the fact that, after the waiting period, the victim reloads the data it previously flushed.
However, when the victim accesses the target address, the victim incurs the cache miss,
while the attacker then observes low latency of the subsequent reload, i.e., a cache hit.
This is why the waiting period is very critical in the \FR{} technique, providing the victim a large enough window to induce the cache miss.

In contrast, the \MCHAMMER{} technique will be the inverse of that logic.
Without any victim access, the default behavior of the \MCHAMMER{} technique will be flushing a cache line that is not actively being executed,
hence it will not trigger a machine clear, and thus the latency of the \texttt{clflush} instruction should be low.
However, if the victim is executing code at the target address, this will trigger a machine clear due to SMC detection mechanisms,
and the latency of the \texttt{clflush} instruction should be significantly higher.

\begin{figure}
    \centering
    \includegraphics[width=\linewidth]{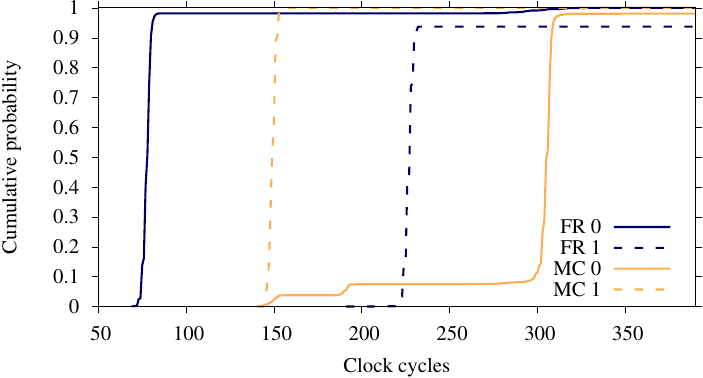}
    \caption{CDFs of histograms for empirical timing data from the \FR{} and \MCHAMMER{} techniques.}%
    \label{fig:latencies_cdf}
  \end{figure}

To test this hypothesis and quantify the previous terms such as ``high latency''
and ``low latency'', we utilize the empirical timing data from the
\autoref{sec:leak} experiments. Nothing changed in the experiment setup---we
gathered \(2^{20}\) timing measurements (\texttt{rdtsc}-based outputs) using the
\FR{} technique, once with the victim executing \texttt{x64\_\-victim\_\-0}
and again with the victim executing \texttt{x64\_\-victim\_\-1}, with the
\FR{} probe set within \texttt{x64\_\-victim\_\-0} in both cases. We then
gathered the same amount of timing measurements using the \MCHAMMER{} technique,
once with the victim executing \texttt{x64\_\-victim\_\-0} and again with the victim
executing \texttt{x64\_\-victim\_\-1}, with the \MCHAMMER{} probe set within
\texttt{x64\_\-victim\_\-0} in both cases. In total, this is four sets of data, and
\autoref{fig:latencies_cdf} illustrates the Cumulative Distribution Functions
(CDF) of the histograms.

We begin the discussion with the results when the victim executes
\texttt{x64\_\-victim\_\-0}, while the attacker targets \texttt{x64\_\-victim\_\-0} with
the \FR{} technique (``FR 0''). We expect that the victim incurs a large
number of cache misses in this case, and the attacker observes low latency in
the reload step, corresponding to a cache hit. The data in
\autoref{fig:latencies_cdf} supports this, with a mean latency of 82.14 clock
cycles and standard deviation of 30.48, and a median latency of 78 clock cycles.

We continue the discussion with the results when the victim executes
\texttt{x64\_\-victim\_\-1}, while the attacker targets \texttt{x64\_\-victim\_\-0} with
the \FR{} technique (``FR 1''). We expect that the victim incurs a large
number of cache hits in this case, and the attacker observes high latency in
the reload step, corresponding to a cache miss. The data in
\autoref{fig:latencies_cdf} supports this, with a mean latency of 250.80 clock
cycles and standard deviation of 92.12, and a median latency of 228 clock
cycles. From the attacker perspective, the difference of \(228-78=150\) clock
cycles is indeed a strong signal.

We continue the discussion with the results when the victim executes
\texttt{x64\_\-victim\_\-0}, while the attacker targets \texttt{x64\_\-victim\_\-0} with
the \MCHAMMER{} technique (``MC 0''). We expect that the victim incurs a large
number of machine clears in this case, and the attacker observes high latency
in the cache line flush step, corresponding to the negative pipeline effects due
to a machine clear. The data in \autoref{fig:latencies_cdf} supports this, with
a mean latency of 311.35 clock cycles and standard deviation of 120.67, and a
median latency of 306 clock cycles.

Finally, we discuss the results when the victim executes
\texttt{x64\_\-victim\_\-1}, while the attacker targets \texttt{x64\_\-victim\_\-0} with
the MC\-Hammer technique (``MC 1''). We expect that the victim incurs a negligible
number of machine clears in this case, and the attacker observes low latency in
the cache line flush step, corresponding to the absence of negative pipeline
effects due to a machine clear. The data in \autoref{fig:latencies_cdf} supports
this, with a mean latency of 149.63 clock cycles and standard deviation of 7.48,
and a median latency of 149 clock cycles. From the attacker perspective, the
difference of \(306-149=157\) clock cycles is indeed a strong signal, a
strikingly similar figure to that of the \FR{} technique (150).

\subsection{A Conceptual Covert Channel}

Given the empirical data in \autoref{fig:latencies_cdf}, we now outline an
abstract covert channel enabled by the \MCHAMMER{} gadget. For ease of exposition,
we utilize the strawman victim in \autoref{fig:victim}. Consider a sender Alice,
wanting to send a single bit to a receiver Bob.

\begin{enumerate}
  \item In an initial step, Alice clears the cache.
  \item To send a 0-value bit, Alice executes a single iteration of the loop within \texttt{x64\_\-victim\_\-0}.
  \item Otherwise, to send a 1-value bit, Alice either executes a single iteration of the loop within \texttt{x64\_\-victim\_\-1}, or does nothing.
  \item To receive the bit, Bob executes the \MCHAMMER{} gadget on a target address within the \texttt{x64\_\-victim\_\-0} function.
  \item If the latency of the \texttt{clflush} instruction is high, this
  indicates a machine clear occurred, implying that Alice did indeed execute code
  within the \texttt{x64\_\-victim\_\-0} function, and Bob receives a 0-value bit.
  \item If the latency of the \texttt{clflush} instruction is low, this
  indicates no machine clear occurred, implying that Alice did not execute code
  within the \texttt{x64\_\-victim\_\-0} function, and Bob receives a 1-value bit.
\end{enumerate}

In the context of the empirical data in \autoref{fig:latencies_cdf}, ``low
latency'' here would be around 149 clock cycles, while ``high latency'' would be
around 306 clock cycles. These values correspond to the medians of the \MCHAMMER{}
gadget data in \autoref{fig:latencies_cdf}.

In a real covert channel, Alice and Bob would be sending more than a single bit,
and would face issues such as synchronization, noise, etc., while simultaneously
attempting to maximize the th\-rough\-put of the channel. These would perhaps be
addressed by utilizing multiple code pages, as well as error-correcting code
techniques. Yet for our purposes, this abstract covert channel serves only as a
stepping stone to utilizing the \MCHAMMER{} gadget as a side channel, which we
introduce in the next section.
\section{\MCHAMMER{}: Side Channel}\label{sec:side}

Thus far, there is seemingly little value of the \MCHAMMER{} gadget over the \FR{} technique.
They have similar impacts on the microarchitectural state (\autoref{sec:leak}),
and similar latencies for signaling purposes (\autoref{sec:covert}).
In this section, we establish the advantage of \MCHAMMER{} over the \FR{}
technique in the context of code (not general data). We start by establishing a
specially-crafted victim and presenting both \MCHAMMER{} and \FR{} traces,
then finally quantify the leakage of these traces for a fair comparison.

\subsection{Side-Channel Traces}

The metrics presented in \autoref{sec:leak} and \autoref{sec:covert} fail to
paint a complete picture from the attacker perspective, due to the
characteristics of the victim process. For example, the strawman victim in
\autoref{fig:victim} for those experiments sets \texttt{VICTIM\_\-ITS} to
\(2^{30}\), while both the \FR{} method and the \MCHAMMER{} method only
collect \(2^{20}\) measurements. In the context of those experiments, this is
essentially equivalent to the victim running in an infinite loop---neither the
\FR{} nor \MCHAMMER{} method need to consider the spacial aspects of the
victim iterations. In a real trace-driven attack, the adversary will want
iteration-by-iteration measurements, and missed accesses represent noise in the
trace, which is often unrecoverable.

To rectify said issue, in this section we modify the strawman victim and
side-channel methods in the following ways.

\begin{enumerate}
  \item In the first experiment, the victim runs \texttt{x64\_\-victim\_\-0}, then runs \texttt{x64\_\-victim\_\-0}.
  \item In the second experiment, the victim runs \texttt{x64\_\-victim\_\-0}, then runs \texttt{x64\_\-victim\_\-1}.
  \item We set \texttt{VICTIM\_\-ITS} to \(2^{10}\).
  \item We collect \(2^{14}\) measurements with the \FR{} method, and the same amount for the \MCHAMMER{} method.
\end{enumerate}

The first two modifications will (eventually) allow us to quantify trace
leakage. The third modification sets a target for the optimal number of accesses
detected by the side-channel methods, i.e., there are now a finite number of
loop iterations and the goal is to capture them all. With the much shorter
execution window implied by the previous modification, the last modification
reduces the number of measurements collected by the side-channel methods, i.e.,
shortens the trace length.

\begin{figure}
    \centering
    \includegraphics[width=\linewidth]{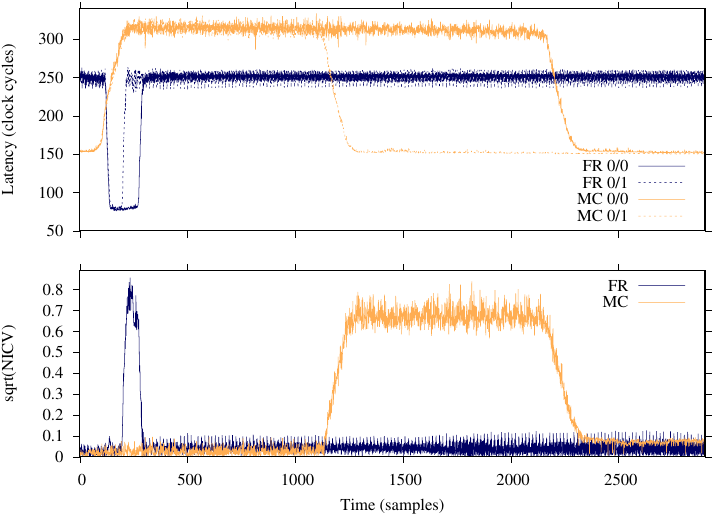}
    \caption{Top: averaged traces for the \FR{} and \MCHAMMER{} techniques, and different
    victim execution paths (i.e.\ classes, 0/0 and 0/1).
    Bottom: the NICV metric's square root, or maximum correlation.
    The plots align and display the same time slice.}%
    \label{fig:nicv}
\end{figure}

We then proceeded to collect four sets of data. The first and second sets are from running the first and second experiments in parallel with the \FR{} technique.
The third and fourth sets are from running the first and second experiments in parallel with the \MCHAMMER{} gadget.
In all of these experiments, both the \FR{} technique and the \MCHAMMER{} gadget use
the same target line as the experiments in \autoref{sec:leak} and \autoref{sec:covert},
i.e., 128 bytes within the \texttt{x64\_\-victim\_\-0} function.
Comparing these traces should give us an idea about spacial aspects of the traces, such as granularity and magnitude of leakage.
We repeated the process a total of 1,000 times, then averaged the traces accordingly.
\autoref{fig:nicv} (Top) depicts the result.

We begin the discussion with the results when the victim executes
\texttt{x64\_\-victim\_\-0} followed by \texttt{x64\_\-victim\_\-1}, while the attacker
targets \texttt{x64\_\-victim\_\-0} with the \FR{} technique (``FR 0/1''). We
expect that the victim experiences cache misses for the first half of this trace,
followed by cache hits for the second half. We can see the data (from the attacker perspective, inverted) is consistent
with this expectation, with low latency measurements (around 70--80 clock
cycles) that span roughly 90 data points starting near the beginning of the
trace. Note this is far below the optimal \(2^{10}\) value we would hope to
detect.

We continue the discussion with the results when the victim executes
\texttt{x64\_\-victim\_\-0} followed by \texttt{x64\_\-victim\_\-0}, while the attacker
targets \texttt{x64\_\-victim\_\-0} with the \FR{} technique (``FR 0/0''). We
expect that the victim experiences cache misses for the complete trace. We can see
the data (from the attacker perspective, inverted) is consistent with this expectation, with low latency measurements
(around 70--80 clock cycles) that span roughly 180 data points starting near the
beginning of the trace. Note this is also far below the optimal \(2^{11}\) value
we would hope to detect.

We continue the discussion with the results when the victim executes
\texttt{x64\_\-victim\_\-0} followed by \texttt{x64\_\-victim\_\-1}, while the attacker
targets \texttt{x64\_\-victim\_\-0} with the \MCHAMMER{} gadget (``MC 0/1''). We expect
that the victim experiences machine clears for the first half of the trace, and
the absence of machine clears for the second half. We can see the data is
consistent with this expectation, with high latency measurements (around
310--320 clock cycles) that span roughly 1,100 data points starting near the
beginning of the trace. Remarkably, note this is comfortably above the optimal
\(2^{10}\) value we would hope to detect.

Finally, we discuss the results when the victim executes
\texttt{x64\_\-victim\_\-0} followed by \texttt{x64\_\-victim\_\-0}, while the attacker
targets \texttt{x64\_\-victim\_\-0} with the \MCHAMMER{} gadget (``MC 0/0''). We expect
that the victim experiences machine clears for the complete trace. We can see the data is
consistent with this expectation, with high latency measurements (around
310--320 clock cycles) that span roughly 2,200 data points starting near the
beginning of the trace. Remarkably, note this is comfortably above the optimal
\(2^{11}\) value we would hope to detect.

To summarize, our data suggests that the waiting period required by the
\FR{} technique severely limits the granularity of resulting traces. In
contrast, the \MCHAMMER{} gadget has no waiting period---it can simply take
repeated measurements immediately, regardless of victim behavior. This results
in increased granularity of \MCHAMMER{} traces compared to those of \FR{}.
Trace granularity is an extremely important aspect of side-channel data
procurement methods, and in fact there is an entire subfield of side-channel
analysis dedicated to improving trace granularity, termed performance
degradation attacks or ``boosting''
\cite{2007:monopoly,2016:degrade,2017:cesar,2022:hd}
We now formalize this difference in trace
granularity for a quantitative comparison.

\subsection{Leakage Compared}

Normalized Inner Class Variance (NICV) is
particularly suited to our needs because it utilizes only public data and is
independent of specific leakage models \cite{2014:NICV}. This characteristic makes NICV an ideal
metric for comparing the quality of trace data \cite[Section~3]{2014:NICV}. NICV, which ranges from 0 to 1,
is mathematically defined as:
\[
\mathrm{NICV}(X,Y) = \frac{\mathrm{Var}[\mathrm{E}[Y | X]]}{\mathrm{Var}[Y]}
\]
where \(Y\) represents the traces, \(X\) represents the classes, and
\(\mathrm{E}\) denotes the expected value or mean. Notably, the square root of
the NICV metric, known as the correlation ratio, serves as an upper bound for
Pearson's correlation coefficient \cite[Corollary~8]{DBLP:journals/tc/ProuffRB09}.

For our purposes, we simplify this metric by considering only two classes (0/0
and 0/1), resulting in the following formula:
\[
\mathrm{NICV}(X,Y) = \frac{(\mathrm{E}[Y | X=0] - \mathrm{E}{[Y|X=1])}^2}{4 \cdot \mathrm{Var}[Y]}
\]
\autoref{fig:nicv} (Bottom) demonstrates the application of this metric to two sets of
measurements, for both the \FR{} and \MCHAMMER{} methods.
It visualizes the square root of NICV, indicating the maximum correlation.
What remains is to interpret the NICV values.

Points of Interest (POIs) are critical in understanding and detecting leakage
within trace data. POIs are specific points in the trace where the detection of
information leakage exceeds a predetermined threshold. This threshold is set to
identify significant deviations that suggest data leakage. When the NICV value
at a certain point in the trace surpasses this threshold, it indicates that the
information related to the variable being measured---in this case, latency---is
leaking, thus revealing sensitive
information about the underlying operations. By marking these points, attackers
can focus on areas with the highest likelihood of leakage, enabling targeted
and efficient side-channel analysis.

By applying a simple threshold to identify POIs, we arrive at the statistics
presented in \autoref{tab:pois}.
A very direct interpretation of this data is as follows.
If fewer than \(2^{10}\) POIs are detected (the number of victim loop iterations), it
suggests a loss of information, meaning the victim is executing faster than the
side-channel technique can measure. In this specific environment, this indicates that
the \FR{} technique cannot provide adequate trace granularity to prevent
information loss, whereas the \MCHAMMER{} gadget achieves this with ease when
selecting a suitable threshold.

\begin{table}[!b]
    \caption{POI counts and ratios at various NICV thresholds with two side-channel techniques (see \autoref{fig:nicv}).
             The ratios (x) are between the POI counts of the two techniques.}%
    \label{tab:pois}
    \centering
        \begin{tabular}{rrr} \hline
          Threshold & \FR{} & \MCHAMMER{} \\ \hline
          0.2 & 86 & 1119 (13.0x) \\
          0.3 & 81 & 1077 (13.3x) \\
          0.4 & 76 & 1030 (13.6x) \\
          0.5 & 67 & 985 (14.7x) \\ \hline
\end{tabular}
\end{table}

In conclusion, the \MCHAMMER{} gadget
results in a statistically significant increase in information
leakage compared to the \FR{} technique, due to a notable increase in POIs
ranging from 13 to 15 fold.
\section{\MCHAMMER{}: Key-Recovery Attack}\label{sec:attack}

Previous sections apply the \MCHAMMER{} gadget to strawman shared libraries.
In this section, we demonstrate that \MCHAMMER{} has practical applications
as a general microarchitectural side-channel technique by carrying out a
cryptanalytic attack against a widely-available cryptographic utility.

\subsection{Digital Signatures and Elliptic Curves}\label{sec:eccbak}

We focus on the signing process of the standardized and widely-deployed
Elliptic Curve Digital Signature Algorithm (ECDSA).
The parameters include a generator point \( G \) on an elliptic curve group of
prime order \( n \), and an approved hash function \( h \) such as SHA-256 or
SHA-512.

To generate a key pair, the private key \( \alpha \) is a randomly selected
integer from the range \( 1 \) to \( n-1 \), inclusive. The corresponding public
key \( D \) is \( [\alpha]G \), where \( [i]G \) is the scalar multiplication
operation.

When Alice wishes to send a digitally signed message \( m \) to Bob, she follows
these steps using her private and public key pair \( (\alpha_A, D_A) \):

\begin{enumerate}
  \item Select a secret nonce \(k\) such that \(0 < k < n\).
  \item Compute \(r = {([k]G)}_x \bmod n\).
  \item Compute \(s = k^{-1} (h(m) + \alpha_A r) \bmod n\).
  \item Alice sends \((m, r, s)\) to Bob.
\end{enumerate}

Note that the \(s\) component of the signature is one equation with two
unknowns, \(\alpha_A\) and \(k\), hence disclosing \(k\) immediately reveals the
private key.
This is a fact we use later in \autoref{ecc:attack}.

\PARAGRAPH{ECDSA leakage}
With side-channel attacks on ECC, typical targets during ECDSA signing are
(i) the scalar multiplication \([k]G\) during the computation of \(r\) \cite{DBLP:journals/csur/LouZJZ21};
(ii) the modular inversion \(k^{-1}\) during the computation of \(s\) \cite{2017:cesar};
(iii) various modular multiplications during the computation of \(s\) \cite{DBLP:journals/tches/Ryan19}.
In this regard, we will focus on (i) due to the nature of the vulnerability we discover and describe later in this section.

There are generally two popular approaches to utilizing leakage from the ECDSA nonce \(k\).
(i) If only part of the nonce leaks,
the attacker queries multiple signatures,
gathers leakage for several nonces,
then combines this information using lattice methods to recover the ECDSA
private key \cite{DBLP:journals/joc/NguyenS02,DBLP:journals/dcc/NguyenS03,DBLP:journals/dcc/Howgrave-GrahamS01}.
These lattice methods have steadily improved over the years to reduce the number of signatures required by
the attacker \cite{DBLP:conf/ches/BengerPSY14,DBLP:conf/ctrsa/PolSY15}.
(ii) If (almost) all of the nonce leaks,
the attacker only needs a single signature and side-channel trace.
For example, \citet[Section 4.3]{DBLP:conf/ccs/WangCPZWBTG17} target leaky scalar multiplication
in \texttt{libgcrypt} to recover the full secret scalar from a single signing,
and similarly several recent works
\cite[Section 6]{DBLP:conf/ndss/GrasGKBR20}
\cite[Section 5.1]{DBLP:conf/uss/PaccagnellaLF21}
\cite[Section 4.6]{DBLP:conf/raid/JiangSK22}
\cite[Figure 4]{2023:dramaqueen}
\cite[Figure 3]{DBLP:conf/date/Wang0XW24}.
In this regard, we will focus on (ii) due to the nature of the vulnerability we discover and describe later in this section.

\subsection{The SECCURE Toolset}

The Secure Elliptic Curve Crypto Utility for Reliable Encryption
(SECCURE)\footnote{\url{http://point-at-infinity.org/seccure/}} is a
lightweight tool for ECC, designed to provide an easy-to-use command-line
interface for cryptographic operations. It features
key generation (the \texttt{seccure\--key} command-line utility),
encryption/decryption (the \texttt{seccure-encrypt} and \texttt{seccure\--decrypt} command-line utilities),
digital signatures (the \texttt{seccure\--sign} and \texttt{seccure\--verify} command-line utilities),
and Diffie-Hellman key exchange (the \texttt{seccure\--dh} command-line utility).
SECCURE supports various standardized elliptic curves and is optimized for
efficiency in constrained environments like embedded systems and mobile devices.
The first release of SECCURE was in July 2006, before side-channel attacks
received much attention. The latest release of SECCURE is v0.5 in August 2014,
and we use this version in our experiments, unmodified.
Packaging wise, SECCURE implements its own elliptic curve arithmetic by
dynamically linking against \texttt{libgcrypt} for multiprecision integer
arithmetic functionality. SECCURE is part of Debian and Ubuntu Linux
distributions.

\subsection{A Single-Trace Key-Recovery Attack}\label{ecc:attack}

\begin{figure}
    \centering
    \includegraphics[width=\linewidth]{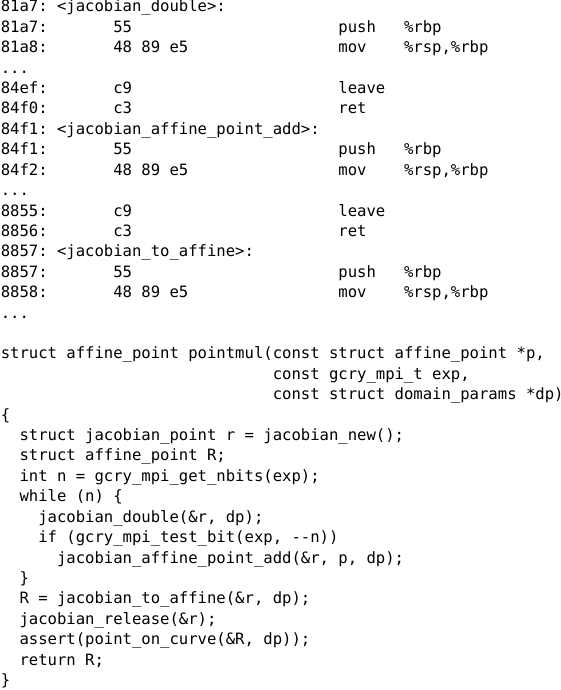}
    \caption{SECCURE's implementation of scalar multiplication.
             The conditional \texttt{jacobian\_\-affine\_\-point\_\-add} is a side-channel vulnerability.}%
    \label{fig:pointmul}
\end{figure}

\PARAGRAPH{A timing attack vulnerability}
\autoref{fig:pointmul} depicts SECCURE's implementation of scalar
multiplication, in the function \texttt{pointmul}. It is a classical
left-to-right binary double-and-add algorithm, in this case using projective
coordinates for point doubling (function \texttt{jacobian\_\-double}) and point
addition (function \texttt{jacobian\_\-affine\_\-point\_\-add}).
This is a fairly textbook example of a side-channel vulnerability, since the
point addition only occurs when scalar bits are set;
see \cite[Section 4.1]{DBLP:journals/csur/LouZJZ21} for an overview.
At a high level,
this algorithm is in fact the same that \citet[Section 8]{DBLP:conf/uss/GrasRBG18} exploit with TLBleed,
yet an implementation within \texttt{libgcrypt} and with a slightly different
elliptic curve cryptosystem (EdDSA \cite{DBLP:journals/jce/BernsteinDLSY12}).

\begin{figure}
    \centering
    \includegraphics[width=\linewidth]{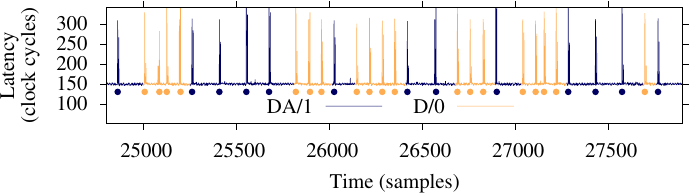}
    \caption{An \MCHAMMER{} partial trace targeting the \texttt{jacobian\_\-double} function calls from \texttt{pointmul} during an ECDSA sign operation.
             Wider gaps between the peaks indicate a point doubling followed by a point addition (DA) and a 1-bit, while narrower gaps indicate a sole point doubling (D) and a 0-bit.
             The 32 dots for the 32 peaks in this excerpt annotate said operations accordingly.
             From left-to-right, the bits are 1\-0\-0\-0\-0\-1\-1\-1\-1\-0\-0\-0\-1\-0\-0\-0\-0\-1\-1\-0\-0\-0\-1\-0\-0\-0\-0\-1\-1\-1\-0\-1.}%
    \label{fig:ecc}
\end{figure}

\PARAGRAPH{Applying \MCHAMMER{}}
The goal of the attacker in this case is to recover the sequence of elliptic curve point doubles and additions.
This information directly reveals the binary representation of the scalar---a single-trace attack.
To implement the attack with \MCHAMMER{}, we \texttt{mmap} the
\texttt{seccure\--sign} utility, and placed an \MCHAMMER{} probe in the middle of the
\texttt{jacobian\_\-double} function.
We then executed the (stock, unmodified) \texttt{seccure\--sign} utility,
specifying the NIST standard P-256 curve, and ran the \MCHAMMER{} gadget
in parallel\@. \autoref{fig:ecc} depicts an arbitrary portion of the trace.
Threat model-wise, targeting the \texttt{seccure\--sign} command-line utility in this way
is no different than, for example, targeting the \texttt{openssl-dgst} command-line utility
for digital signatures, a long-established methodology \cite{DBLP:conf/asiacrypt/BrumleyH09}.

The trace excerpt in \autoref{fig:ecc} contains 32 peaks where the latency exceeds 250 clock cycles.
These peaks are the \texttt{pointmul} function processing 32 bits of the (secret) ECDSA scalar \(k\),
where each peak delimits a single iteration of said function's \texttt{for} loop.
Contiguous peaks that are separated by roughly 70 samples correspond to a clear bit and a single double operation (D), while those separated
by roughly 120--150 samples correspond to a set bit and a point doubling coupled with a point addition operation (DA).
Gleaning this full D and DA sequence directly reveals the full binary representation of the ECDSA nonce \(k\),
and we solve for the ECDSA private key \(\alpha_A\) using the linear equation for the ECDSA signature component \(s\) in \autoref{sec:eccbak}.

To conclude, this successful key-recover attack---against
unmodified code in a distribution-provided command-line utility,
and only a single signature, i.e., one cryptographic operation---demonstrates
the efficacy of \MCHAMMER{},
in particular, its ability to track secret-dependent code traversal.
\section{Related Work}\label{sec:related}

\PARAGRAPH{Machine clears for performance degradation}
\citet{2022:hd} introduce HyperDegrade, a novel technique that combines previous
approaches to performance degradation with simultaneous multithreading (SMT)
architectures. It investigates the root causes of performance degradation using
cache eviction, largely attributed to machine clears. The slowdown produced by
HyperDegrade is significantly higher than previous approaches, leading to an
increased time granularity for \FR{} attacks.

\PARAGRAPH{Constructing transient execution gadgets via machine clears}
\citet{2021:rage} investigate the root causes of a subset of transient execution
attacks, specifically focusing on machine clears. The authors identify previously
unexplored types of machine clears induced by different functionality such as
floating point operations, self-modifying code detection, memory ordering, and
memory disambiguation. These machine clear events create new transient execution
windows, widening the horizon for known attacks and introducing new attack
gadgets.

\PARAGRAPH{Cache flushing as a microarchitectural timing attack vector}
The Flush+\-Flush technique by \citet{2016:flfl}
exploits the \texttt{clflush} instruction's
execution time variance due to early abort behavior.
The key insight of the Flush+\-Flush attack is that the execution time of the
\texttt{clflush} instruction varies depending on whether the data is cached or
not. If the data is not cached, the execution time is shorter; if the data is
cached, it takes longer. While \MCHAMMER{} shares some similarity to the
Flush+\-Flush technique, the root causes are significantly different:
(i) According to the authors \cite[Section~3]{2016:flfl},
      ``[Flush+\-Flush] builds upon the observation that the \texttt{clflush}
      instruction can abort early in case of a cache miss'',
      while the \MCHAMMER{} root cause is linked to machine clears;
(ii) The Flush+Flush technique yields timing differences between 9--12 clock
      cycles \cite[Figure~1]{2016:flfl}, while \MCHAMMER{} yields timing differences
      of roughly 157 clock cycles (see \autoref{fig:latencies_cdf}),
      a huge improvement in terms of trace granularity and error-handling ability.
\section{Conclusion}\label{sec:conclusion}

In this study, we introduced \MCHAMMER{}, a novel microarchitectural side-channel
technique that exploits machine clears to extract sensitive information. Unlike
most traditional side-channel methods, \MCHAMMER{} operates without memory access or
waiting periods, making it highly efficient.

\MCHAMMER{} outperforms traditional side-channel methods like \FR{}.
We quantified this improvement using the NICV leakage metric,
indicating a 13--15 fold increase in trace granularity.

By tracking machine clears through the latency of the \texttt{clflush}
instruction, \MCHAMMER{} provides insights into secret-dependent code traversal.
As such, we applied \MCHAMMER{} to successfully recover private keys from an open
source, unmodified, deployed public key cryptography implementation using only a single trace.
Along these lines, mitigations for \MCHAMMER{} are similar to those for \FR{}
and are well-understood at the software level---using
constant-time software engineering practices,
where all code and data addresses are assumed to be public in the threat model.

As side-channel techniques evolve and improve, understanding novel methods like
\MCHAMMER{} becomes crucial for securing software and hardware. By thoroughly
examining the root cause of microarchitectural effects and casting a wide net
during analysis, we learn increasingly more about the interaction between CPU
components at the microarchitectural level.

In closing, we are pleased to contribute \MCHAMMER{} as yet another tool in the
belt of the side-channel research community.
 
\bibliographystyle{ACM-Reference-Format}

\end{document}